\documentclass[twocolumn]{aastex63} 

\hypersetup{linkcolor=red,citecolor=blue,filecolor=cyan,urlcolor=blue}

\usepackage{graphicx,color}
\usepackage{amssymb}
\usepackage{amsmath}
\usepackage{url}
\usepackage{natbib}
\usepackage{txfonts}

\usepackage{url}
\usepackage{hyperref}

\begin{document}
	
	\title{Are the brightest coronal loops always rooted in mixed-polarity magnetic flux?}
	
	\author[0000-0001-7817-2978]{Sanjiv K. Tiwari}
	\affiliation{Lockheed Martin Solar and Astrophysics Laboratory, 3251 Hanover Street, Bldg. 252, Palo Alto, CA 94304, USA}
	\affiliation{Bay Area Environmental Research Institute, NASA Research Park, Moffett Field, CA 94035, USA}
	
	\author[0000-0002-6478-3281]{Caroline L. Evans}
	\affiliation{Department of Physics, Davidson College, 
		Box 6910, Davidson, NC 28035, USA}
	\affiliation{Department of Physics, North Carolina State University, Raleigh, NC 27695, USA}
	
	\author[0000-0001-7620-362X]{Navdeep K. Panesar}
	\affiliation{Lockheed Martin Solar and Astrophysics Laboratory, 3251 Hanover Street, Bldg. 252, Palo Alto, CA 94304, USA}
	\affiliation{Bay Area Environmental Research Institute, NASA Research Park, Moffett Field, CA 94035, USA}
	
	\author[0000-0003-0819-464X]{Avijeet Prasad}
	\affiliation{Center for Space and Aeronomic Research, The University of Alabama in Huntsville,
		320 Sparkman Drive,
		Huntsville, AL 35805, USA}
	
	\author[0000-0002-5691-6152]{Ronald L. Moore}
	\affiliation{NASA Marshall Space Flight Center, Mail Code ST 13, Huntsville, AL 35812, USA}
	\affiliation{Center for Space and Aeronomic Research, The University of Alabama in Huntsville,
		320 Sparkman Drive,
		Huntsville, AL 35805, USA}

	\begin{abstract}
		
	 A recent study demonstrated that freedom of convection and strength of magnetic field in the photospheric feet of active-region (AR) coronal loops, together, can engender or quench heating in them. Other studies stress that magnetic flux cancellation at the loop-feet potentially drives heating in loops. We follow 24-hour movies of a bipolar AR, using EUV images from SDO/AIA and line-of-sight (LOS) magnetograms from SDO/HMI, to examine magnetic polarities at the feet of 23 of the brightest coronal loops. We derived FeXVIII emission (hot-94) images (using the Warren et al. method) to select the hottest/brightest loops, and confirm their footpoint locations via non-force-free field extrapolations. From 6"$\times$6" boxes centered at each loop foot in LOS magnetograms we find that $\sim$40\% of  the loops have both feet in unipolar flux, and $\sim$60\% of the loops have at least one foot in mixed-polarity flux. The loops with both feet unipolar are $\sim$15\% shorter lived on average than the loops having mixed-polarity foot-point flux, but their peak-intensity averages are equal. The presence of mixed-polarity magnetic flux in at least one foot of majority of the loops suggests that flux cancellation at the footpoints may drive most of the heating. But, the absence of mixed-polarity magnetic flux (to the detection limit of HMI) in $\sim$40\% of the loops suggests that flux cancellation may not be necessary to drive heating in coronal loops -- magnetoconvection and field strength at both loop feet possibly drive much of the heating, even in the cases where a loop foot presents mixed-polarity magnetic flux.
	\end{abstract}
	

	\section{Introduction} \label{sec:intro}
	
		Magnetic energy dissipated in coronal loops by unknown processes heats the Sun's corona to millions of Kelvin. The brightest and hottest extreme ultraviolet (EUV) and X-ray solar coronal loops are rooted in strong magnetic flux in active regions (ARs) \citep{golu80,fish98,dahl18,ugar19,asga19}. These loops have temperatures of 2-6 MK, or more. The processes for heating them to these temperatures remain ill-determined \citep{zirk93,schr98,moor99,asch05,kats05,klim06,real14,hino19}. The two most well-known mechanisms that could explain these temperatures are magentohydrodynamic (MHD) waves \citep[e.g.,][and references therein]{vanb11} and nanoflare heating \citep{parker72,parker83b,parker88}.  

	In both cases, magnetoconvection most probably drives the magnetic energy input \citep[e.g.,][]{tiw17}. Photospheric convection can produce MHD waves that transport energy to higher parts of the Sun's atmosphere \citep{prie94,prie02}. Photospheric convective motion can also randomly shuffle the feet of the coronal loops so that they become entangled and braided, dissipating the magnetic energy by current sheet dissipation in the higher solar atmosphere \citep{parker83b, parker88}. Recent observations of an AR and modelling show evidence of braided magnetic structures in the corona \citep{cirt13, thal14,tiw14,pont17}.
	
	Some studies \citep{falc97,tiw14, tiw17, tiw19,chit17,chit18, prie18} find the presence of mixed-polarity magnetic flux at the feet of the brightest coronal loops and suggest a third manner of driving heating -- by flux cancellation at the loop feet. According  to these studies, the brightest AR coronal loops most frequently have at least one footpoint in a region of mixed-polarity magnetic flux. This implies that over time, magnetoconvection causes an increase in the injection of free magnetic energy into the brightest coronal loops via some consequence of the opposite-polarity flux, which is most probably magnetic flux cancellation, often accompanied by small-scale magnetic flux emergence \citep{tiw19,sahi19}.

	Magnetic reconnection events taking place very low in the chromosphere, accompanied by magnetic flux cancellation, evidenced by fine-scale explosive events and chromospheric inverted-Y-shaped jets in the lower solar atmosphere at these sites, can feed energy and hot plasma into the corona \citep{chit17, tiw19,pane19,pane20}.   Magnetic flux cancellation is most probably the result of submergence of lower reconnected loops \citep[e.g.,][and references therein]{tiw19}. \cite{prie18} have described a theoretical model of how chromospheric and coronal heating of loops might depend on flux cancellation speed, flux size, and field strength in the loop \cite[see also][]{synt20}. An observational test supporting this model was recently performed by \cite{park20}.   
	
	\cite{tiw17} demonstrated that photospheric magnetic rooting plays an important role in determining the amount of heating in AR coronal loops -- freedom of convection and strength of magnetic field in the loop-feet, together, can enhance or suppress heating in coronal loops. Using EUV observations and non-linear force-free  modelling of two ARs they found that the hottest loops of an AR are the ones connecting sunspot umbra/penumbra at one end to (a) penumbra, (b) unipolar plage, or (c) mixed-polarity plage on the other end. The loops connecting dark sunspot umbra at both ends were not visible in EUV images. Thus, these loops are the coolest loops, despite being rooted in the strongest magnetic field regions. They concluded that both the field strength and freedom of convection at the loop feet play crucial role in determining the heating magnitude of the loop. 
	As mentioned earlier, some recent investigations stress more on the loop-foot mixed-polarity (above-mentioned connectivity `c'), suggesting that flux cancellation is involved in heating chromospheric and coronal loops. In the present work we investigate whether all or most of the hottest loops of an AR have mixed-polarity magnetic flux at their feet. If not, what percentage of them are rooted in unipolar magnetic flux at each end of the loop? 
	
	If it turns out that at least one foot of each hot loop has mixed-polarity magnetic flux, then it would provide strong evidence to the idea of flux cancellation being involved in driving heating of coronal loops. The presence of unipolar field at both feet of hot loops will support the idea that (irrespective of polarity mixture at the loop feet) heating of the hottest coronal loops depends primarily on the freedom of convection at the loop feet, together with the strength of the magnetic field there \citep{tiw17}, not primarily on flux cancellation. 

	\begin{figure*}
		\centering
		\includegraphics[trim=0cm 11cm 0cm 7cm,clip,width=\linewidth]{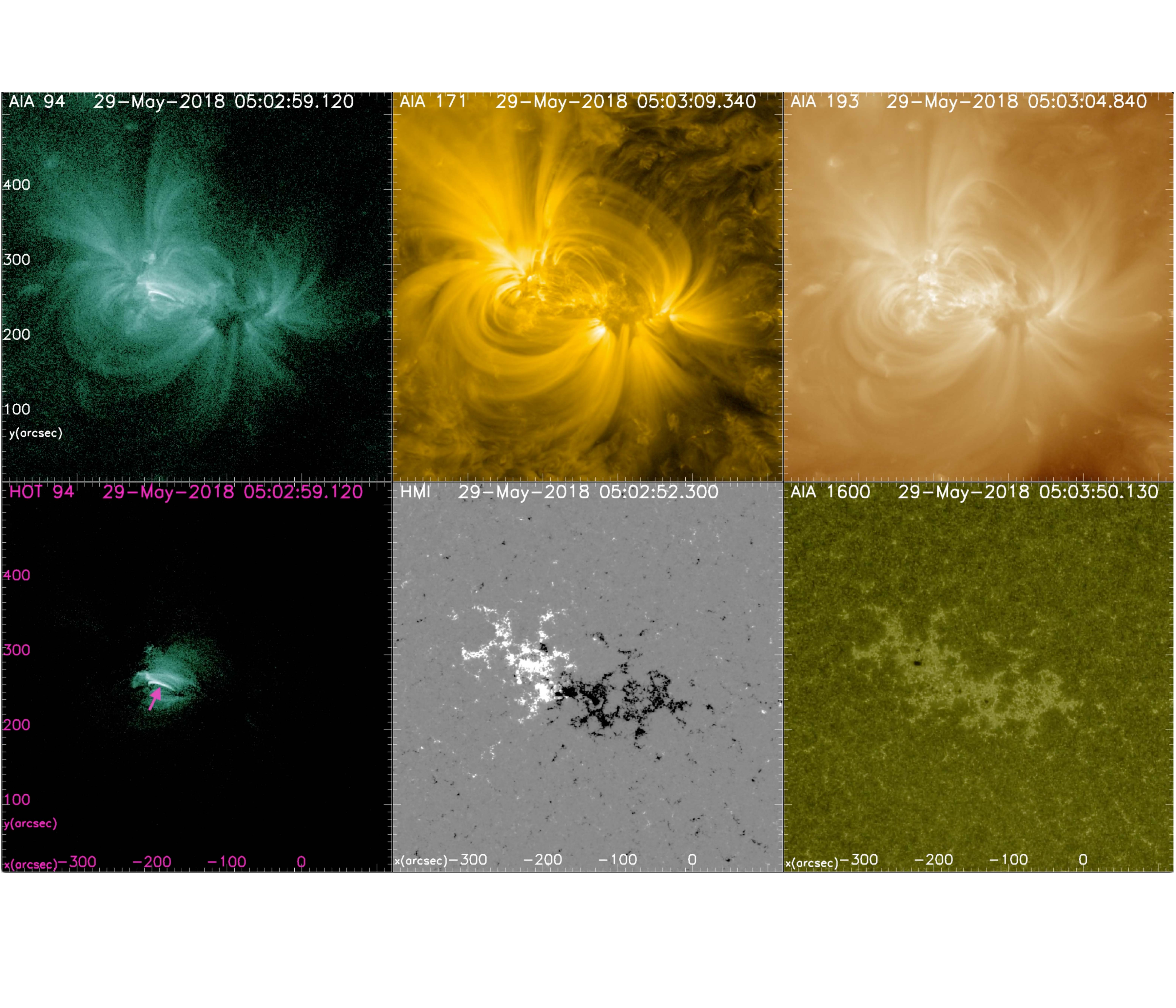}
		\caption{Context images of the NOAA AR 12712 on 29-May-2018 at about 05:03 UT. The six image panels contain six different wavelengths --  \emph{Top Left:} 94 \AA, \emph{Top Center:} 171 \AA, \emph{Top Right:} 193 \AA, \emph{Bottom Left:} our created hot 94, \emph{Bottom Center:} HMI LOS magnetogram, \emph{Bottom Right:} 1600 \AA. An arrow in hot 94 panel points to a loop that is listed in Table \ref{tab:1} as Loop 6, and is shown in Figures \ref{fig:2_loops} and \ref{fig:3_nff}. This figure is an image frame from the movie ``Movie1.mp4". The full animation for 24 hours at a three-minute temporal cadence is available online (Movie1.mp4).} 
		\label{fig:1_context}
	\end{figure*}

	\section{Data and Modelling}\label{sec:data}

	We examine EUV/UV images of NOAA AR 12712 obtained with Atmospheric Imaging Assembly \citep[AIA:][]{leme12} on-board Solar Dynamics Observatory \citep[SDO][]{pesn12} to investigate the hot emissions centered on the Fe XVIII line (6-8 MK), by treating the data to omit the 1 MK plasma detection. To isolate the brightest and hottest coronal loops of the AR in question, we use the method laid out in \cite{warr12} to subtract out the warm component of the 94 \AA\ intensity:
	
	\begin{equation}
	I_{94warm} = 0.39\sum_{i=1}^{4} a_i \left[\frac{f I_{171} + (1-f) I_{193}}{116.54}\right]^i ,
	\end{equation}
	where I$_{171}$ and I$_{193}$ are the respective intensities of AIA 171 \AA\ and AIA 193 \AA; f is determined to be 0.31; $a_i$ are, in order, $-7.31\times10^{-2}, 9.75\times10^{-1}, 9.90\times10^{-2},$ and $2.84\times10^{-3}$.
	
We refer to Fe XVIII emission images calculated by the above method as hot 94 images. To create the hot 94 images, we downloaded 94 \AA, 171 \AA, and 193 \AA\ AIA data at a three-minute cadence for the 24 hours of May 29, 2018 from Joint Science Operations Center (JSOC) with two im\_patch parameters: (1) a center at -230", 270" and (2) a box height and width of 600" (Figure \ref{fig:1_context}). The EUV channels have a 12-second temporal cadence and a 1.2" resolution (0.6" pixel size) \citep{leme12}. However a 3-minute cadence worked well for our purpose because most of the hottest loops lived well beyond 3-minutes, the shortest one living for about 12 minutes (see Table \ref{tab:1}).

The 94 \AA\ channel captures the characteristic emission of Fe XVIII from plasma at 6-8 MK, but also captures emission of plasma around 1 MK \citep{warr12}. The 171 \AA\ channel detects the characteristic emission of the Fe IX line from plasma at about 0.8 MK and the 193 \AA\ channel shows the characteristic emission of Fe XII from plasma at around 1.5 MK \citep{real11,leme12}. All images were normalized by dividing each image by its exposure time.

For investigating photospheric magnetic flux polarity at the loop feet, we downloaded line-of-sight (LOS) magnetograms from the Helioseismic and Magnetic Imager \citep[HMI:][]{scho12, sche12}, also onboard SDO, of the same field of view (FOV) as AIA EUV images at a 3-minute cadence, the same cadence as used for AIA images. 
In Figure \ref{fig:1_context}, we show our active region NOAA 12712 in UV, EUV, processed AIA images and a processed HMI magnetogram. This AR is of interest because it is a bipolar region observed close to the solar disk center during an otherwise quiet Sun. This AR was also observed on this day (May 29, 2018) by Hi-C 2.1 \citep{rach19}. 

All our generated maps (AIA and HMI) were processed and de-rotated using SolarSoft routines \citep{free98}. 
We discarded the one time frame in this set that shows very large noise. We examined the UV AIA data at 1600 \AA\ to confirm alignment between each of the EUV wavelengths and LOS magnetograms. After these data treatments we followed each hot loop in our 24-hour span of observations to select the most clearly visible hot loops, which are sufficiently isolated from other hot loops in the surroundings. This led to selection of 23 hot loops in the 24 hours of data. For each loop, we made a light curve of the emission in a 2" x 2" box placed on the brightest segment of the loop top to obtain the loop's start, peak-brightness, and end times.
	
Note that in a few cases there are two loops tangled in the way that they have a single foot on one end. Because these loops are spatially isolated from other bright loops, we included them in our study and counted these as two separate loops.

\begin{figure*}
		\centering
		\includegraphics[trim=21cm 10cm 22cm 6cm,clip,width=0.91\linewidth]{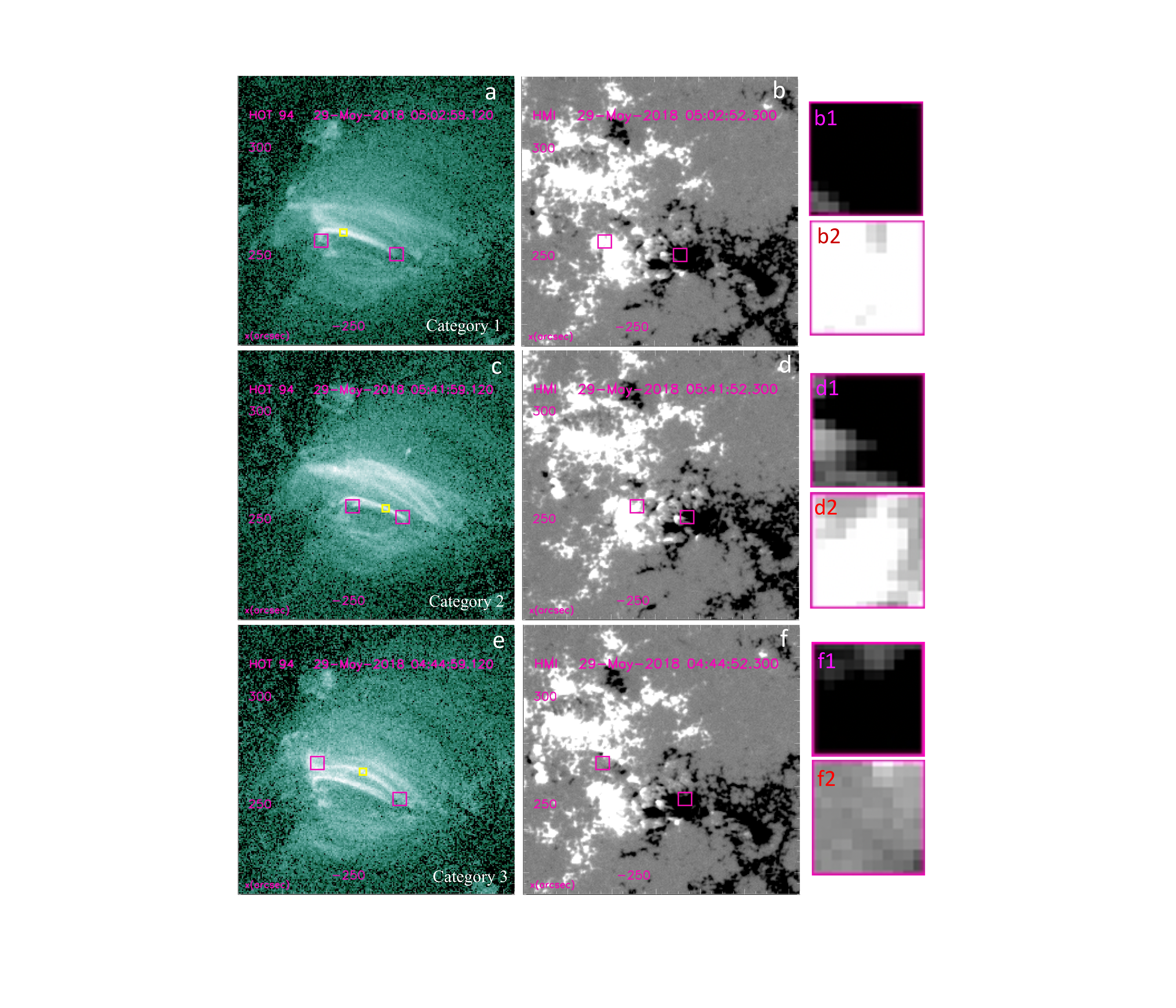}
		\caption{Three example loops depicting the three alternative categories. The left column (panels a, c, e) presents a close view of the loops (from Movie1.mp4) in hot 94 and the right column (panels b, d, f) presents the same FOV of LOS magnetograms. Insets b1, b2, d1, d2 and  f1, f2 are each a further zoomed-in view of the LOS magnetogram of each loop foot. White/black/grey colours in the LOS magnetograms are for positive/negative/zero field. Category 1 (uppermost row) has two unipolar feet -- shown example is Loop 6 in Table \ref{tab:1}; Category 2 (middle row) has two mixed-polarity feet -- shown example is Loop 8 in Table \ref{tab:1}; Category 3 (bottom row) has one unipolar and one mixed-polarity foot -- shown example is Loop 4 in Table \ref{tab:1}. Pink boxes on the hot 94 images and on the LOS magnetograms outline the 6" x 6" area examined in the LOS magnetograms and give the corresponding field-strength and polarity histograms (see Fig. \ref{fig:4_histo}). Yellow boxes on the loop outline the 2" x 2" area integrated over to obtain the peak intensities listed in Table \ref{tab:1}.}
		\label{fig:2_loops}
	\end{figure*}

After visual identification of the footpoints of the hottest coronal loops using a zoomed-in FOV, we chose the locations and placed boxes (of size  6''$\times$6'') centered at the loop feet (see pink boxes in Figure \ref{fig:2_loops}). To confirm the placement of these boxes we performed coronal magnetic field extrapolations from photospheric vector magnetograms. The coronal magnetic field extrapolations were from HMI vector magnetograms \citep[Space weather HMI Active Region Patches -- HMI SHARPs:][]{bobr14} using the non-force-free extrapolation technique described in \cite{hu10}. The visualizations were carried out in VAPOR software \citep{li19}. The extrapolation technique works best when the bottom boundary is flux-balanced and the field strength is significantly higher than the magnetogram noise level. 

Because the AR remained close to disk center (within 30 degrees from the disk center) during our observation period, we did not deproject the magnetograms for our loop extrapolations. Deprojection is usually not required if the AR is within 30 degrees from the solar disk center \citep[e.g.,][]{falc16}.

 Coordinates and times had to be approximated between the HMI SHARPs and HMI LOS Magnetograms due to their different temporal cadences of 12 and 3 minutes (used here), respectively. To accommodate this approximation, we examined the three LOS magnetograms before and after the identified peak time of each loop to ensure no significant flux changes occur. To find the coordinates of each of the two feet of each loop, we first made sure that the observed loop closely traced its model field lines and then selected the coordinates of each foot to be those of that foot of the model field lines.


For each of the footpoints of the 23 loops, we made a histogram of the LOS magnetic field strength and polarity from the 144 pixels inside the 6" x 6" box centered on the footpoint to determine which loops had two unipolar footpoint boxes, which had one unipolar box and one mixed-polarity box, and which had mixed-polarity flux in both footpoint boxes.

\begin{figure}
		\centering
		\includegraphics[trim=0cm 0cm 0cm 0cm,clip,width=0.91\linewidth]{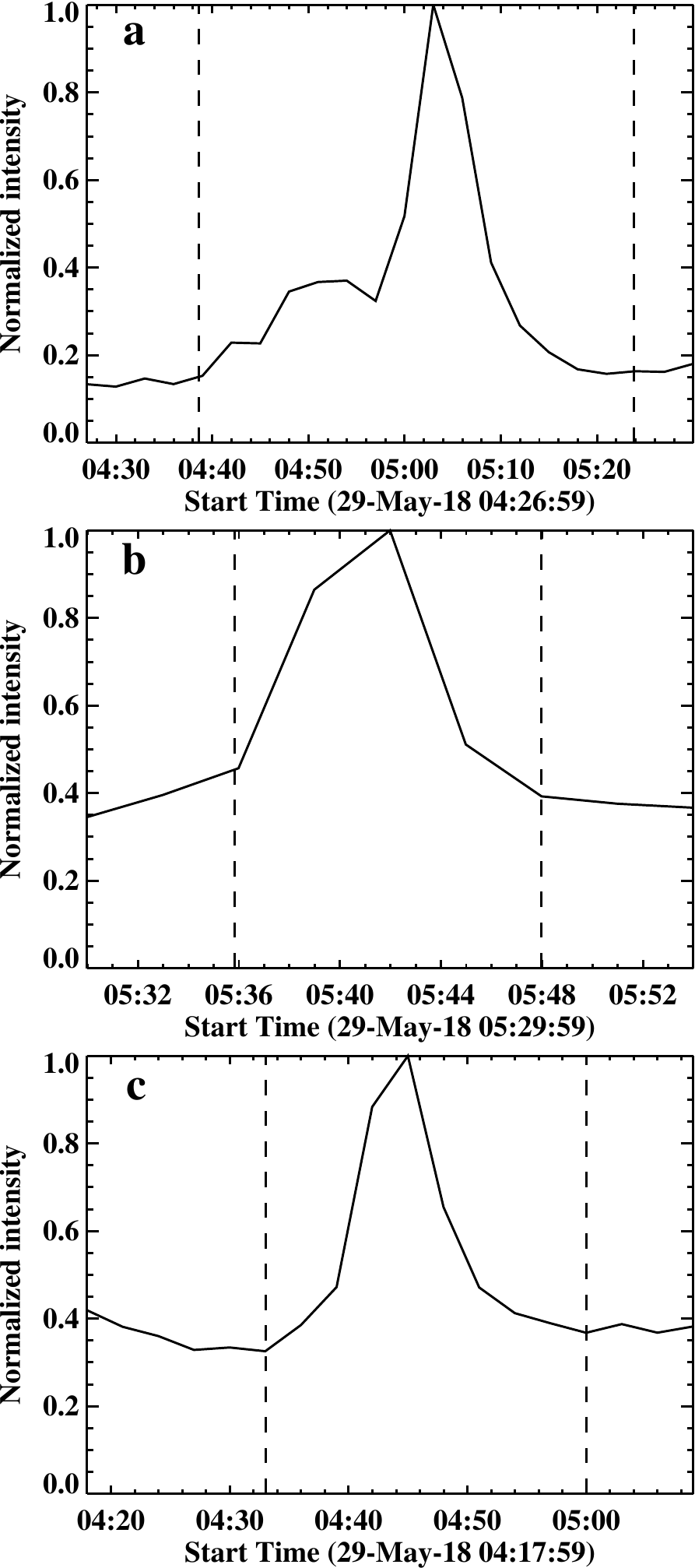}
		\caption {Panels a, b, and c show normalized intensity curves (light curves of hot 94) integrated over the area  inside the yellow box for the three loops shown in Figure \ref{fig:2_loops}a, c and e, respectively. The dashed vertical lines in each panel mark the start and end times of the loop (also verified with the visual inspection of each loop).}  
		\label{fig:lightcurves}
	\end{figure}

We also measured peak intensities of all loops. Based on the size of the thinnest loops in our sample, we selected a box size of 2" by 2" to measure the peak intensity of the loop, see the yellow boxes in Figure \ref{fig:2_loops}. For this, we integrated intensity of the hot 94 image inside the 2"$\times$2" box during the peak time of each loop. The box was placed at several places along the loop to find out the maximum value of the integrated intensity. 
	For each loop, we used the light curve of the 2" x 2" box to find the start time, the peak-brightness time, and the end time of the loop, and to obtain the intensity of the emission in the box at the peak-brightness time (the loop's peak intensity given in Table \ref{tab:1}). The light curves for three loops are shown in Figure \ref{fig:lightcurves}. 



\begin{deluxetable*}{cccccccc}
	\tablenum{1}
	\tablecaption{23 Selected Hot Coronal Loops from AR 12712 on May 29, 2018}
	\tablewidth{0pt}
	\renewcommand{\arraystretch}{1.0}

	\tablehead{\colhead{Loop} & \colhead{Start Time} & \colhead{Peak Time} & \colhead{End Time} & \colhead{Lifetime\tablenotemark{a}} &\colhead{Footpoint 1} & \colhead{Footpoint 2} & \colhead{Peak Intensity } \\
		\colhead{Index} & \colhead{UT} & \colhead{UT} & \colhead{UT} & \colhead{Minutes} &\colhead{Coord\tablenotemark{b} \& Polarity\tablenotemark{c} } & \colhead{Coord\tablenotemark{b} \& Polarity\tablenotemark{c} } & \colhead{DN s$^{-1}$}
	}
	
	\startdata
	1  & $01:53:59$ & $02:05:59$ & $02:41:59$ & $48$ & $(-255, 243)$ $  mix$& \bf{(-281, 246)} $  +$ & $1764$ \\
	2  & $02:14:59$ & $02:35:59$ & $03:11:59$ & $57$ & \bf{(-240, 251)} $  -$& \bf{(-280, 274)}   $  +$ & $1098$ \\	
	3  & $03:59:59$ & $04:20:59$ & $04:44:59$ & $45$ & $(-242, 237)$ $  mix$& \bf{(-260, 242)}  $  +$ & $1684$ \\	
	4  & $04:33:59$ & $04:44:59$ & $04:59:59$ &  $27$  & \bf{(-227, 254)} $  -$& $(-264, 271)$ $  mix$  & $872$ \\
	5  & $04:29:59$ & $04:44:59$ & $05:17:59$ & $48$ & \bf{(-230, 251)} $  -$& \bf{(-265, 259)} $  +$ & $1215$ \\
	6  & $04:38:59$ & $05:02:59$ & $05:23:59$ & $45$ & \bf{(-228, 252)} $  -$ & \bf{(-262, 258)} $  +$ & $2055$ \\
	7  & $05:38:59$ & $05:41:59$ & $05:50:59$ & $12$ & \bf{(-223, 251)} $  -$& \bf{(-255, 258)} $  +$ & $1257$ \\
	8  & $05:35:59$ & $05:41:59$ & $05:47:59$ & $12$ & $(-226, 253)$ $  mix$ & $(-248, 258)$ $  mix$ & $862$ \\
	9  & $05:41:59$ & $05:53:59$ & $06:26:59$ & $45$ & $(-226, 245)$ $  mix$& \bf{(-252, 247)}  $  +$ & $995$ \\	
	10  & $05:56:59$ & $06:35:59$ & $07:35:59$ & $99$ & \bf{(-216, 253)} $  -$& $(-243, 260)$ $  mix$ & $847$ \\
	11  & $06:20:59$ & $06:47:59$ & $07:38:59$ & $78$ & \bf{(-212, 251)} $  -$& \bf{(-243, 259)}  $  +$ & $1144$ \\
	12  & $06:20:59$ & $06:47:59$ & $07:38:59$ & $78$ & $(-216, 255)$ $  mix$& \bf{(-239, 257)}  $  +$ & $904$ \\
	13  & $07:35:59$ & $07:56:59$ & $08:20:59$ & $45$ & \bf{(-202, 252)} $  -$& $(-236, 262)$  $  mix$ & $742$ \\
	14  & $07:35:59$ & $07:56:59$ & $08:20:59$ & $45$ & $(-203, 253)$ $  mix$& \bf{(-230, 258)}  $  +$ & $559$ \\
	15  & $09:35:59$ & $09:44:59$ & $09:53:59$ & $18$ & $(-189, 238)$ $  mix$& \bf{(-215, 243)}  $  +$ & $3245$ \\
	16  & $09:56:59$ & $10:08:59$ & $10:20:59$ & $24$ & $(-186, 239)$ $  mix$& \bf{(-207, 242)}  $  +$ & $2096$ \\
	17  & $10:56:59$ & $11:08:59$ & $11:32:59$ & $36$ & \bf{(-171, 251)} $  -$& \bf{(-209, 259)}  $  +$ & $955$ \\
	18  & $10:56:59$ & $11:08:59$ & $11:32:59$ & $36$ & $(-172, 258)$ $  mix$& \bf{(-204, 271)}  $  +$ & $646$ \\
	19  & $12:29:59$ & $12:38:59$ & $13:14:59$ & $45$ & $(-156, 242)$ $  mix$& \bf{(-195, 249)}  $  +$ & $1323$ \\
	20  & $12:35:59$ & $12:53:59$ & $13:14:59$ & $39$ & \bf{(-150, 252)} $  -$& \bf{(-190, 273)}  $  +$ & $1447$ \\
	21  & $12:35:59$ & $12:53:59$ & $13:14:59$ & $39$ & \bf{(-150, 251)} $  -$& \bf{(-195, 259)}  $  +$ & $827$ \\
	22  & $17:05:59$ & $17:23:59$ & $17:32:59$ & $27$ & \bf{(-121, 243)} $  -$& \bf{(-149, 250)} $  +$ & $986$ \\
	23\tablenotemark{d}  & $21:38:59$ & $22:44:59$ & $23:38:59$ & $120$ & $(-71, 240)$ $  mix$& \bf{(-96, 245)} $  +$ & $573$ \\
	\hline
	average & -- & -- & -- & 46$\pm$6\tablenotemark{e} & -- & -- & 1221$\pm$129\tablenotemark{f}
	\enddata	
	\tablenotetext{a}{The uncertainty in the measurement of lifetime of a coronal loop can be up to six minutes, twice the 3-min cadence of AIA data used for the presented analysis.}
	\tablenotetext{b}{Coordinates of the center of the box outlining the footpoint.}
	\tablenotetext{c}{Footpoint 1 of each loop is rooted in dominantly negative magnetic polarity flux region. Footpoint 2 of each loop is rooted in dominantly positive magnetic polarity flux region.}
	 \tablenotetext{d}{This loop has the most prolonged heating (displaying several sequential pulses) of our 23 loops. }
	\tablenotetext{e}{The mean lifetime of these 23 loops is 46 min and the standard deviation of that mean is $\pm$6 min. Average lifetime for the loops having at least one mix-polarity foot is 49$\pm$8 min, and that for the loops having each foot in unipolar flux is 42$\pm$6 min. Thus, the lifetime of the loops with both feet unipolar are marginally significantly ($\sim$15\%) shorter lived.} 
	\tablenotetext{f}{Average peak intensity for the loops having mixed-polarity flux at foot one or both feet comes out to be 1222$\pm$200 DN s$^{-1}$, and that for the loops having unipolar flux at both  feet comes out to be 1220$\pm$120 DN s$^{-1}$. This shows that there is insignificant difference in the peak intensities of the loops having mixed-polarity flux or unipolar flux at their feet.}

	\tablecomments{The table contains information for each loop investigated in this study. The given coordinates for each loop foot are for each foot's center on the photosphere. Times of interest (start, peak, end) are given in addition to total lifetime of each loop. Coordinates of unipolar feet are in bold font for easy identification. Peak intensities are the integrated intensity inside a 2"$\times$2" box placed on the brightest region of each loop during its peak brightness time; see yellow boxes in Figure \ref{fig:2_loops} for three examples. All numbers in the table are rounded to their closest integer.}

	\label{tab:1}
\end{deluxetable*}

\begin{figure}
	\centering
	\includegraphics[trim=25cm 2cm 39cm 3cm,clip,width=0.91\linewidth]{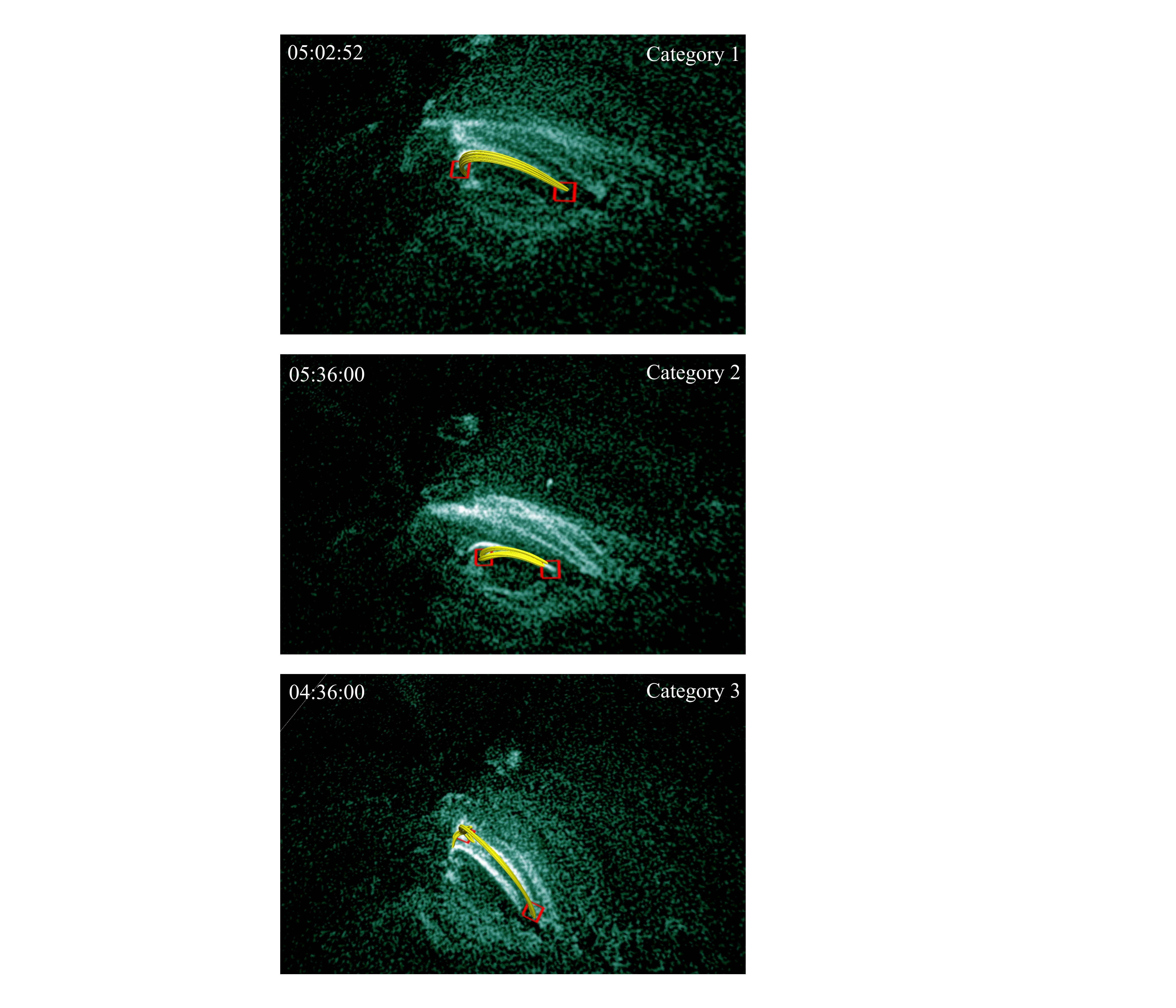}
	\caption{Sample 3D loop reconstructions shown by the VAPOR software. The three panels present each of the three example loops in Figure \ref{fig:2_loops}. Yellow lines represent the extrapolated field lines. Red boxes in each panel are the same as those in Figure \ref{fig:2_loops}. Note that our extrapolated field lines are rotated to the viewing angle of the observed AR.  } 
	\label{fig:3_nff}
\end{figure}

	\begin{figure*}
		\centering
		\includegraphics[trim=15cm 16cm 13cm 3cm,clip,width=\linewidth]{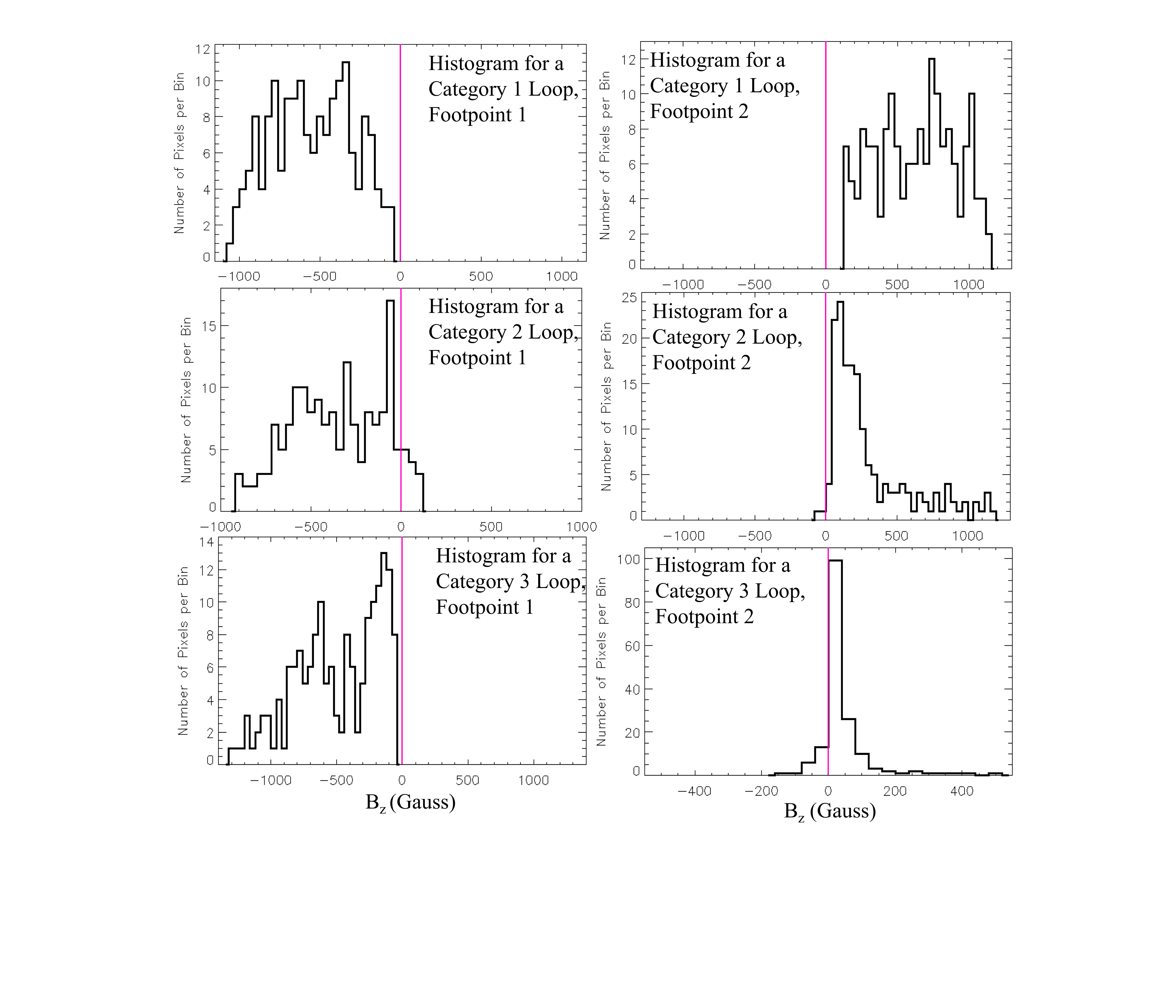}
		\caption{Histograms of the LOS magnetic field strength and polarity at each footpoint of the three example loops displayed in Figure \ref{fig:2_loops}, one loop for each of the three alternative categories. From Table \ref{tab:1}, the example loop for category 1 is Loop 6, the example loop for category 2 is Loop 8, and the example loop for category 3 is Loop 4. The left panels present histograms of footpoints 1 of the three loops (the foot located in the negative majority polarity region) and the right panels present histograms of footpoints 2 of the three loops (the foot located in the positive majority polarity region).  The vertical pink line in each panel marks $B_z$ = 0. }
		\label{fig:4_histo}
	\end{figure*}


	\section{Results} 
	
	We found 23 of the brightest coronal loops that qualify under our selection criterion for loops described in Section \ref{sec:data} (e.g., loops should be bright and hot enough to be clearly visible in hot 94, they should be fairly isolated from other loops in the surroundings, and peak well in light curves). In Table \ref{tab:1}, we list the 23 selected hot coronal loops, three of which are presented in detail in Figures \ref{fig:2_loops}, and \ref{fig:3_nff}. Each of the 23 loops is marked by pink arrows in Movie1.mp4, in three frames (during the peak intensity time, on a frame just before the peak intensity time, and on a frame just after the peak intensity time). Our use of the hot 94 technique ensures that the selected loops are over 1 MK \citep{warr12}.

In Table \ref{tab:1}, we give the start, peak, and end times of each loop found via visual tracking of the loops as well as from light curves, and also give the overall lifetime of each loop. We followed each loop from their peak time in forward and backward directions in time, visually and in light curves, to define the loop start/end time, which is when the loop gets almost invisible (or shows the lowest intensity in light curves) in hot 94 images in backward/forward time from their peak time.
The lifetimes of different hot loops vary from 12 minutes to two hours, with an average lifetime of 46$\pm$6 minutes.

Typical coronal loops have a lifetime of 20--30 minutes \citep{mulu11,pete12,real14}, but the lifetime of the smallest loops can be less than a minute to a few minutes \citep{wine13,tiw19}, and the hottest loops can live upto multiple hours \citep{lope2007,klim10}. Most of the longest lived loops are probably a result of interaction of multiple loop strands, which sequentially heat making the loop bundle live much longer than a loop strand \citep[e.g.,][]{warr02}. Loop 23 in Table \ref{tab:1} shows this behavior of sequential heating (see Movie1.mp4).

	The coordinates of the center of the box outlining the two footpoints for each loop are also given in Table \ref{tab:1}. For easy identification, the coordinates in bold font indicate that the surrounding box has unipolar magnetic flux; all others have mixed-polarity magnetic flux. 
	We quantified the distribution of magnetic field strength and polarity at each foot of each loop via histograms of LOS magnetograms within the selected footpoint region, in the way presented in Figure \ref{fig:4_histo} for three examples.
	We use a 6"$\times$6" boxed area centered at each foot to obtain LOS magnetogram histograms at each foot. The 6"$\times$6" size of the box ensures that the loop foot is completely covered within the box.

	We have three categories of loops: 1. loops having unipolar magnetic flux in both feet, 2. loops having mixed-polarity magnetic flux in both feet, and 3. loops having one foot in mixed-polarity flux and the other foot in unipolar magnetic flux. Of the 23 loops that we examined, $\sim$40\% (9/23) have both feet in unipolar magnetic flux, $\sim$4\% of loops (1/23) have both feet in mixed-polarity flux, and $\sim$56\% (13/23) have one foot in unipolar and one in mixed-polarity flux.  
	
		We have considered the presence of mixed-polarity only when the LOS magnetogram at the loop foot contains values $\ge$ 20 G of positive or negative minority polarity flux. The random noise level in the LOS magnetograms is about 7 G \citep{couv16}, fairly well below our selected lower limit. We would have counted footpoint 2 of our only category 2 loop as unipolar if there was no minority-polarity pixel with a negative $B_z$ value larger than 20 G, i.e., if there were no pixel in the second bin (in 20--40 G range, bin size is 20 G) on the negative/minority polarity side of the zero line in this footpoint box's histogram in Figure \ref{fig:4_histo}. This case is the most marginal one in our sample -- in all other mix-polarity cases there are several to many minority-polarity pixels with the magnitude of $B_z$ larger than 20 G.
	
	Because there is only one loop in the category of both loop-feet having mixed-polarity flux, in our discussions we often count that in the category of loops having at least one foot in mixed-polarity flux region.
	
	 In Figure \ref{fig:2_loops} we show three example loops, one from each of the three categories. In the top row of Figure \ref{fig:2_loops}, we show an example loop that has both feet in unipolar magnetic flux (category 1); the peak time in the hot 94 image comes at 05:02:59 and the associated LOS HMI magnetogram is obtained at 05:02:52. The light curve for this loop in Figure \ref{fig:lightcurves}a also shows another (weaker) peak, at 04:50:59. In such cases we count peak brightness at the time of the brightest intensity peak, e.g., at 05:02:59 in this case.
	  In Figure \ref{fig:3_nff} (top panel), we show the extrapolated loop with the correct perspective of the loop, given by means of VAPOR. Its time is 05:02:52 (top panel in Figure \ref{fig:3_nff}). In Figure \ref{fig:4_histo}, top row, we plot histograms for each footpoint (labelled in the figure) of this loop. 
	
	Figure \ref{fig:2_loops} further shows a loop where both feet have mixed-polarity flux (category 2). The peak-intensity time for this loop is 05:41:59 (the displayed LOS HMI magnetogram has a time of 05:41:52). The middle panel of Figure \ref{fig:3_nff} shows the extrapolated field lines obtained from the HMI SHARP vector magnetogram at 05:36:00. The histograms for the feet of this example loop are plotted in the middle row of Figure \ref{fig:4_histo}. 
	
	The last set of images in Figure \ref{fig:2_loops} show an example loop having one footpoint in unipolar magnetic flux and the other footpoint in mixed-polarity flux (category 3). The peak time in the hot 94 image is 04:44:59 and the closest LOS HMI magnetogram is at 04:44:52. The bottom panel in Figure \ref{fig:3_nff} contains the extrapolated field lines. The bottom row of Figure \ref{fig:4_histo} shows the histograms of the two feet of this loop.

	Our non-force-free field extrapolations match well with the observed loops, confirming the selection of footpoint locations (e.g., in Figure \ref{fig:3_nff}). Histograms of LOS magnetograms clearly show whether a loop  foot has a unipolar, or mixed-polarity magnetic field (e.g., in Figure \ref{fig:4_histo}). All of our results are listed in Table \ref{tab:1}.  
	

	We also measured the peak intensity of each loop inside a 2"$\times$2" box during its peak intensity time (see yellow boxes in Figure \ref{fig:2_loops}) and list the values in Table \ref{tab:1}. The peak intensity time for each loop was selected based on visual inspection of loops in the hot 94 movie. The location of the area for calculating peak intensity was also decided via visual inspection of the loops. To make sure the selected area was on the brightest location of the loop, we placed our 2"$\times$2" box at several places, as needed, along each loop.  There is no significant difference in the peak intensities of two major categories of loops (i) having unipolar flux at both of their feet or (ii) having at least one foot in mixed-polarity magnetic flux (see Table \ref{tab:1}, comment {\it `e'}).

	\section{Discussion and Conclusions}
	
	We performed our observational analysis and modelling efforts to investigate if magnetic flux cancellation (inferred from the presence of mixed-polarity magnetic flux) at the loop feet could be a significant heating mechanism for the hottest and brightest loops in the solar corona of the NOAA AR 12712. 
	
    Our finding that 60\% of the loops (14 out of 23 loops under investigation) contain HMI-detected mixed-polarity magnetic flux at least at one of their feet is consistent with the idea of magnetic flux cancellation being involved in heating these loops. The magnetic flux cancellation at a footpoint plausibly resulted from magnetic reconnection in the lower solar atmosphere of coronal loops releasing stored magnetic energy \citep{tiw14,chit18,tiw19}. Recent examinations of magnetic flux cancellation, such as that of \cite{chit18}, note that only part of the dissipating magnetic energy reaches into the corona -- some reaches only into the chromosphere. 
   
   However, the absence of HMI-detected mixed-polarity magnetic flux from the feet of about 40\% of the loops investigated here challenges this idea and allows the possibility that magnetoconvection, in tandem with the magnetic field strength at the loop footpoints, could alone be responsible for loop heating \citep{tiw17}. The same mechanism might dominate in heating the loops having mixed-polarity magnetic flux at one or both of their feet. 

	We did not visually notice any significant magnetic flux emergence and/or cancellation near the loop-feet during lifetimes of any loop. Furthermore, we could not establish magnetic flux emergence and/or cancellation at the feet of the loops over time because of not being able to track the adjacent opposite-polarity flux adequately for this purpose. This reason is also described in \cite{tiw19} for several of the loops found in the core of the same AR. 
	
	The process of loop selection benefited significantly by the use of hot 94 emission. Most of the times hotter loops were very cleanly isolated in hot 94 images and were not so clearly isolated in AIA 94, 193 or 171 \AA\ images. Many loops were clearly identifiable in the hot 94 images, but not in 94 \AA, 193 \AA, or in 171 \AA\ images.  We selected the loops partially also by our ability to perform non-force-free loop extrapolations -- we were forced to throw out the cases in which the VAPOR software presented too many small loops making it difficult to isolate the loop of interest. 
	
	The AR investigated is at the peak of its lifetime during the 24-hour movie, and starts decaying during or immediately after the observations we use in our study. Thus, it is suitable for investigations such as those presented here, avoiding effects of pervasive flux emergence (found in the early phase of ARs) or obvious cancellation (found in the decaying phase of ARs).  
	
	The peak intensities of the loops with unipolar flux at both feet versus the loops with mixed-polarity flux at least at one foot do not show a significant difference, thus suggesting that polarity mixture at a loop-foot (or at both feet) probably does not provide additional heating to the loops. On the other hand, the loops having no HMI-detected mixed-polarity flux at either foot had marginally significantly shorter lifetime than the loops having some HMI-detected mixed-polarity flux at one foot or both feet. This suggests that shorter-lived below-HMI-detectability mixed-polarity flux might have been present at the apparently unipolar feet of these loops and might have been the main driver of the shorter-lived coronal heating in these loops. We note however that the brightness of the loops depends on other factors such as loop length, and area expansion with height \citep[e.g.,][]{klim06,wine08,real14,dahl18,hino19}, which have not been taken into account. Thus, the loop-heating problem requires extensive further investigation.

	In the AR investigated here, there are no fully developed sunspots, as compared to the ARs studied in \cite{tiw17}. Therefore the loops selected here have mostly plage-to-plage connections, and no sunspot connections as were described in \cite{tiw17}. We have therefore studied here coronal loops having plage-to-plage connections (only one class of those described in the above-mentioned study), and explored what percentage of such loops have mixed-polarity flux at their base and what percentage are unipolar at both feet.
	We note that this percentage depends heavily on the selection criterion of loops. Thus, our results have limited absolute significance and cannot be extended to all loops, not even to hot ones only.
	
	There are two possibilities from the present study: either (i) future studies using new generation telescopes giving higher spatial resolution and higher sensitivity magnetograms will confirm that the presence of mixed-polarity magnetic flux, thus flux cancellation, is universal for coronal loop heating, or (ii) the heating is mainly from unipolar flux and it does not matter much whether the loop feet contains a mixed-polarity flux or not —mainly field strength and convective freedom at the loop feet determine how much the loop is heated. 
	
	Our loop extrapolations serve to confirm where the footpoints are for each of the loops investigated. Given the cadence difference between the HMI LOS and HMI SHARP magnetograms, some error might result. However, given that the loop lifetimes are usually $\ge$ 24 minutes, the small difference in time between the HMI LOS and vector magnetograms is probably negligible. The excellent visual agreement between the loops of interest as seen on the hot 94 images and the extrapolated loops viewed with VAPOR from the correct orientation calculated from the outward normal at the patch center gives us a high degree of confidence in our selection of the boxes surrounding the footpoints. 
	
	Further, the size of the box of 6" by 6" probably includes the footpoint and little more. The selected foot area is slightly larger to make sure any part of the loop foot is not missed for the histograms. Thus, it is possible that we counted a few pixels of surrounding area not in the foot of a loop. As a result we might have over-estimated the number of loops with a foot (or both feet) in mixed-polarity. Thus, the number of loops with a mixed-polarity foot in our study can be considered to be at the upper limit, while the ones with both feet unipolar can be considered to be at their lower limit. This would then result into a larger number of loops with unipolar flux at both of their feet, thus providing further strength to the idea that only magnetoconvection, together with the strength of magnetic field at the loop feet (irrespective of the presence/absence of mixed-polarity magnetic flux), drives most of the loop heating. In this scenario both pictures -- MHD waves and nanoflares, can contribute significantly to the bright-loop heating in AR 12712.          
	
	Further, \cite{real19} found that hot spots in the transition region are the footpoints of very hot and transient coronal loops, which often show strong magnetic interactions and rearrangements. Thus, they concluded that hot bright loops often result from magnetic tangling and presumably by large angle reconnection, see also \cite{test14,test20a} and \cite{test20b}. A similar scenario is possible at least in a few  of our loops that are entangled (see, e.g., loops peaking at 05:41:59, 06:47:59, and 12:53:59 UT).
	
	Two main limitations of the present study are 1. limited sample of loops, and 2. limited spatial resolution of the HMI LOS magnetograms. We also do not know if these results based on one AR's loops are valid for other larger and more complicated ARs. Similar isolated ARs with proximity to disk center would be suitable candidates for further investigation. Future research using bigger samples of loops from different ARs and better magnetogram data, e.g., from the Daniel K. Inouye Solar Telescope \citep{trit16}, should validate or challenge the present results. 


We thank the referee for constructive comments. S.K.T., N.K.P. and R.L.M acknowledge the support from NASA HGI program. S.K.T. gratefully acknowledges support by NASA contract NNM07AA01C (Hinode). C.L.E. was supported by funding from NSF grant AGS - 1460767 for the UAH/MSFC Heliophysics REU. She thanks A. Sterling and D. Falconer for their useful discussion during her REU project in Huntsville (during May-August 2019). N.K.P's research was supported by NASA grant NNG04EA00C (SDO/AIA). A.P. acknowledges partial support of NASA grant 80NSSC17K0016 and NSF awards AGS-1650854  and AGS-2020703. AIA and HMI are instruments onboard the Solar Dynamics Observatory, a mission for NASA's Living With a Star program. The AIA and HMI data are courtesy of NASA/SDO and the AIA and HMI science teams. We acknowledge use of the visualization software VAPOR (http://www.vapor.ucar.edu) to generate relevant graphics. This research has made use of NASA's Astrophysics Data System and of IDL SolarSoft package.


\end{document}